\documentstyle [twocolumn,prb,aps] {revtex}
\begin{document}
\draft
\begin{title}
{Molecular dynamics in shape space and 
femtosecond vibrational \\ spectroscopy of  
metal clusters} 
\end{title} 
\author{Constantine Yannouleas and Uzi Landman }
\address{
School of Physics, Georgia Institute of Technology,
Atlanta, Georgia 30332-0430 }
\date{}

\maketitle
\begin{abstract}
We introduce a method of molecular dynamics in shape space aimed at metal
clusters. 
The ionic degrees of freedom are described via a dynamically 
deformable jellium with inertia parameters derived from an incompressible,
irrotational flow. The shell correction method is used to calculate the
electronic potential energy surface underlying the dynamics.
Our finite-temperature simulations of Ag$_{14}$ and its ions, following the
negative-to-neutral-to-positive scheme, demonstrate the potential
of pump-and-probe ultrashort laser pulses as a spectroscopy of 
cluster-shape vibrations. 
\end{abstract}

\pacs{ } 

\narrowtext

The study of the motion of an incompressible and irrotational fluid mass with 
an ellipsoidal surface has its origins in investigations on the theory of the
shape of the Earth. Such studies started with Newton and Maclaurin and were 
continued by others, in particular for the case of a {\it varying\/} surface
by Dirichlet, Dedekind, and Riemann (see Ref.\ \onlinecite{lamb}). 
In these early studies in the realm of astronomy and astrophysics, the 
forces were due to the gravitational field. More recently, similar ideas
were used in nuclear physics in investigations of low-energy collective
isoscalar modes \cite{bm} (as well as of the fission process \cite{nix}
in atomic nuclei). 
Additionally, many features of the analysis of these nuclear 
collective modes \cite{bm} go back to the work of Rayleigh on the normal
modes of a classical liquid drop. \cite{rayl}

In this letter, we introduce and apply a method of molecular dynamics in
shape space (MDSS) aimed at studies of materials clusters, with the 
potential energy surfaces (PES's), which underlie the dynamics, determined via
electronic structure calculations using the shell correction method (SCM),
\cite{yann0,yann1,yann2,yann3} and with inertia parameters specified
according to the aforementioned theory of an irrotational fluid mass.

In the past several years, first-principles molecular-dynamics (FPMD) 
simulation methods, where the dynamics of the ions is evaluated on
concurrently calculated electronic PES's [e.g., through
local density functional (LDF) theory], have been introduced and employed in 
investigations of atomic and molecular clusters. \cite{land,andr} The MDSS 
method complements such techniques. Rather than simulating the
dynamical trajectories of the individual ions as in other FPMD methods, the
MDSS focuses on the dynamics of {\it collective shape variations\/} 
of clusters \cite{note11} (with the ionic degrees of freedom described 
via a dynamically deformable jellium), allowing for reliable and economical 
simulations in a broad cluster size range.

Over the past several years, the development and application of fast
spectroscopical techniques opened new avenues for probing dynamical
processes in matter (both in the molecular and condensed-phase regimes) with
unprecedented temporal resolution (in the femtosecond range
\cite{zewa}).
As an illustration, we apply the MDSS to an
investigation of the shape dynamics of a small neutral cluster (Ag$_{14}$),
motivated by recent experiments using pump-and-probe ultrashort laser pulses.
In these experiments, referred to as NeNePo, \cite{wost}
negative-to-neutral-to-positive, mass selected neutral clusters are prepared 
by electron vertical detachment from the associated negative ions, 
and their subsequent time evolution is probed by photoionization, which 
can be time-delayed with reference to the initial detachment.
Our finite-temperature MDSS simulations of Ag$_{14}$ and of
associated ions, following the NeNePo sequence, demonstrate the potential of
such experiments as a spectroscopy of cluster shape vibrations.
As the jellium description of metal clusters is known to be most adequate
above the Debye temperature, $\Theta _D$, we demonstrate the MDSS method for 
cluster shape dynamics at $T > \Theta_D$(Ag) (= 215 K).

Modelling the dynamics of the ellipsoidal jellium background as that of an 
irrotational, incompressible fluid, its kinetic energy is given \cite{lamb} by
\begin{equation}
{\cal T} (a,b) = \frac{2}{15} \pi \rho a^\prime b^\prime c^\prime
( \dot{a}^{\prime\;2} + \dot{b}^{\prime\;2} + \dot{c}^{\prime\;2} )~,
\end{equation}
where $\rho$ is the mass density of the fluid, and $a^\prime$,
$b^\prime$, and $c^\prime$ are the principal semi-axes of the ellipsoid.
Since the PES $V(a^\prime, b^\prime)$ depends only on two semi-axes,
we can eliminate the third semi-axis 
$c^\prime$ from the kinetic energy by using the volume 
conservation, namely  
$a^\prime b^\prime c^\prime = R^3$. By further defining reduced semi-axes
$a=a^\prime/R$, $b=b^\prime/R$, and $c=c^\prime/R$, we can rewrite the
kinetic energy as,
\begin{equation}
{\cal T} (a,b) = \frac{1}{2} M_{aa} \dot{a}^2 +
M_{ab} \dot{a} \dot{b} +
\frac{1}{2} M_{bb} \dot{b}^2~, 
\end{equation}
where the elements of the inertial-mass matrix $M$ are 
$M_{aa}=C (1+1/a^4b^2)$,
$M_{ab}=M_{ba}=C/a^3b^3$, and $M_{bb}=C (1+1/a^2b^4)$, with
$C=4\pi R^5 \rho /15$. 

At constant energy,
the motion of the shape will be governed by Hamilton's equations associated
with the hamiltonian $H={\cal T}(a,b) + V(a,b)$. Introducing new symbols
$a \leftrightarrow q_1$, $b \leftrightarrow q_2$, $p_a \leftrightarrow p_1$,
and $p_b \leftrightarrow p_2$, for the generalized coordinates $a$ and $b$ and
for the associated canonical momenta $p_a$ and $p_b$, Hamilton's equations
of motion are given by
\begin{eqnarray}
\dot{q}_i &=& \sum_{j=1}^2 (M^{-1})_{ij} p_j~,  \\
\dot{p}_i &=& -\frac{\partial V}{\partial q_i}
-\frac{1}{2} \sum_{j,k=1}^2 
\frac{\partial (M^{-1})_{jk}}{\partial q_i} p_jp_k~,
\label{eq4}
\end{eqnarray}
where $M^{-1}$ is the inverse of the inertial-mass matrix $M$.
These equations are solved using a 6th-order Runge-Kutta method.

MDSS simulations can be carried out at constant energy or constant 
temperature. The latter (canonical-ensemble) simulations can be achieved 
through a straightforward modification of Hamilton's equations according to the
Nos\'{e}-Hoover-thermostat prescription, \cite{noho} namely by adding a 
variable friction term $-p_i \eta/Q$ to the right-hand-side of the $\dot{p}_i$ 
equation, where $Q$ is a constant characterizing the reservoir and 
$\dot{\eta}=2{\cal T}(q_1,q_2)-2k_{\text{B}}T$, with $T$ the temperature and 
$k_{\text{B}}$ the Boltzmann constant. Since, however, the triaxial MDSS 
involves two independent variables only, we use a chain \cite{mart} 
of three Nos\'{e}-Hoover thermostats in order to guarantee that the 
equilibrium ensemble will exhibit canonically distributed positions and 
momenta.   

The SCM approach has been derived by us originally from the LDF theory
(for details, see Refs.\ \onlinecite{yann0,yann2}). This method, as well
as a semiempirical version \cite{yann1,yann2} of it which we employ here
to determine the PES's [i.e., $V$ in Eq.\ (\ref{eq4})], has been shown to 
yield accurate results \cite{yann0,yann1,yann2,yann3,note2} when compared to 
experiments and/or self-consistent Kohn-Sham LDF calculations (when available) 
for several metal-cluster systems. 
Particularly pertinent to this study are SCM
results for the vertical Ionization Potential (vIP) of Ag$_{14}$ 
and the electron vertical Detachment Energy (vDE) of Ag$_{14}^-$; the
SCM results, calculated for the equilibrium configurations, are vIP=5.84 eV 
and vDE=2.20 eV, in excellent agreement with measured values, vIP=5.90 eV 
\cite{alam} and vDE $\sim$ 2.2 eV. \cite{tayl}

We prepare first an equilibrium ensemble of 10$^3$ phase-space points 
[$(q_1)_0$, $(q_2)_0$, $(p_1)_0$, and $(p_2)_0$] of the anion (Ag$_{14}^-$)
by selecting them from a long trajectory (up to 400 ps) 
on the PES of Ag$_{14}^-$ generated using the Nos\'{e}-Hoover dynamics
at a given temperature $T$. These phase-space
points are subsequently used as initial values for generating 
shorter (up to 3 ps), constant-energy trajectories, \cite{note6} 
with an integration time step of 0.5 fs, \cite{note4} on
the PES of the neutral Ag$_{14}$ (which is produced from the anion by 
electron photodetachment caused by the first laser pulse). At each time step,
we determine the classical probability, ${\cal P} (t)$, that 
vertical ionization of the dynamically evolving neutral cluster is 
energetically allowed through 
the two-photon absorption process induced by the probe-laser pulse.
${\cal P} (t)$ is that  fraction of the instantaneous shape configurations 
(averaged over the 10$^3$ constant-energy trajectories) with a
vertical ionization potential vIP$\leq 2\hbar \omega$, where 
$\hbar \omega$ is the single-photon energy.

The probability ${\cal P}(t)$ is a classical quantity which in principle is 
not the sole factor determining the yield of the 
photoionization process (e.g., the photoion current of the
cationic daughters). Indeed the photoion current is proportional to the product
of the classical ${\cal P}(t)$ with the quantum mechanical, vibrational,
Frank-Condon overlap factor. \cite{herz} If the neutral clusters happen to be
produced in shape configurations far away from the PES minima of the 
positive ion, the Franck-Condon overlap factor can be so low initially that 
virtually 
no positive ions are generated, regardless of the value of ${\cal P}(t)$,
and thus the onset of the photoion current can exhibit a characteristic delay.
The NeNePo experiment for the Silver trimer conforms to this case.
\cite{wost,note3}
However, the case of Ag$_{14}$ is different, since here the PES minima of the 
neutral and the positive clusters overlap substantially (see Fig.\ 1). 
As a result, in the latter case, the Franck-Condon factor for the ionization
process is close to 
unity, and we can safely assume that the photoion current is mainly 
proportional to the classical probability ${\cal P}(t)$.

In Fig.\ 1, we  display the PES's for Ag$_{14}^-$ (bottom panel), Ag$_{14}$
(middle), and Ag$_{14}^+$ (top). For the neutral
Ag$_{14}$, which is axially symmetric, there are two energetically distinct
isomers -- an oblate (O) one and a prolate (P) one. On the associated PES
(Fig.\ 1, middle panel), there are six local minima, three
(energetically degenerate ones) 
corresponding to the oblate O-isomer (having two equal axes larger than 
unity;
the case $a=b > 1$ with $c=1/a^2$ is easily identifiable; the other two 
follow by circular permutation of the indices $a$, $b$, and $c$, i.e.,
pseudorotations) and the other three (energetically degenerate ones)
to the P-isomer (having two equal axes smaller than unity; 
the case of $a=b < 1$ with $c=1/a^2$ is also easily identifiable,
and the other two are obtained via pseudorotations). 
The positive and negative ions
are slightly triaxial, and as a result their PES's exhibit 12 local minima,
six of them close to oblate shapes and the other six close to prolate
shapes; in each case, the six minima are energetically degenerate and are
related to each other via pseudorotations.

At $T=300$ K, the canonical-ensemble distribution of the initial 10$^3$ 
phase-space points 
on the PES of Ag$_{14}^-$ is distributed about the O-like
minima, since for the anion these minima are the global ones
by $\approx 0.06$ eV (or 700 K) compared to the P-like minima. 
As a result of the vertical photodetachment, some of the initial shapes will
land on the basins of the P minimum of the neutral Ag$_{14}$;
nevertheless most of the initial shapes will land on the basins of
the O minimum. For the neutral Ag$_{14}$, these latter basins are deep,
separated from the basins of the P minimum by a barrier
of $\approx$ 0.35 eV (i.e., 4060 K), and thus only initial 
O-shapes associated with momenta far in the tails of the canonical distribution
will be able to make excursions outside these basins. 
Consequently, the majority 
of the (constant-energy) trajectories of Ag$_{14}$ 
will correspond to shape 
vibrations around the O minimum (see the Lissajous-type trajectory
in the middle panel of Fig.\ 1).

To investigate how these vibrations can be probed by a 
photoionizing laser pulse of a given frequency, 
we display in Fig.\ 2 the vIP surface [which is 
the difference between the PES of the cation and that of the neutral
Ag$_{14}$ (see Fig.\ 1)]. It is noteworthy that the local minima
of the neutral Ag$_{14}$ do not correspond to extrema in the vIP's. Indeed
the O minimum has a vIP of 5.85 eV which is intermediate between 
the extremal values of 6.0 eV and 5.7 eV.
Consequently, 
a photoionizing laser pulse with photon energy 
$2 \hbar \omega=$ 5.85 eV 
is ideal for probing the vibrations 
about the O minimum, since the ionization condition vIP 
$\leq 2 \hbar \omega$ will be satisfied only for certain segments of 
the constant-energy trajectories, and thus the measured photoionization yield
will exhibit oscillations reflecting the shape vibrations. 

An inspection of the probability ${\cal P} (t)$ (which is displayed
in Fig.\ 3) corroborates indeed the above analysis,
and in addition allows for determination of the period of the 
shape vibrations around the O minimum of Ag$_{14}$.
For $T=300$ K (lower dashed-dotted curve labelled 1), 
the probability ${\cal P} (t)$
exhibits large amplitude oscillations with a period of $\approx$ 800 fs.
Raising the temperature of the initial Ag$_{14}^-$ ensemble 
tends to progressively smear out the
oscillatory structure in ${\cal P}(t)$. This behavior is a consequence of
the fact that the distribution of the initial phase-space points 
on the PES of the anion is more
spread away from the region of the O-like minima (a relatively
larger amount of
initial points is distributed over the region of the P-like minima).
As a result, a larger, but still not dominant, fraction of constant-energy 
trajectories on the PES of Ag$_{14}$ is restricted to the regions
inside the basins of the P minima. For $T=600$ K (lower dotted curve
labelled 2), 
the effect of such trajectories is to
reduce the amplitude of the oscillations of the ${\cal P}(t)$ curve.
We also note here that for the range of temperatures considered (which covers
the experimentally expected range) no influence on the period of the 
oscillations is predicted. 

With a somewhat larger probe-laser frequency 
(e.g., $2 \hbar \omega=$ 5.90 eV), 
the ionization condition vIP $ \leq 2 \hbar \omega$ is satisfied for 
longer segments of the constant-energy trajectories, and as a result the 
troughs in the ${\cal P} (t)$ curve are now less accentuated compared to the
case of $2 \hbar \omega=$ 5.85 eV (see upper curves labelled 3 and 4 in
Fig.\ 3).

We remark that the experiments suggested in our paper, and the NeNePo 
experiments performed in Ref.\ \onlinecite{wost}, 
involve electron detachment and ionization of clusters, and they can be
performed with photon energies which are away from the plasmon 
excitations of silver clusters, which otherwise would have complicated
the interpretation of such experiments due to resonant photoabsorption.
\cite{note1} 

In summary, a molecular dynamics in shape space
method for simulating the dynamics of low-frequency 
collective modes associated with metal-cluster shapes was introduced.
These collective modes (adiabatic in the Born-Oppenheimer sense) are analogous
to the well known low-energy isoscalar vibrations of atomic 
nuclei. \cite{bm} The ionic degrees of freedom were described via a 
dynamically deformable jellium with inertia parameters derived from an 
incompressible, irrotational flow, and the shell correction method was used to 
calculate the electronic potential energy surface underlying the dynamics
of the cluster shape.
Our finite-temperature simulations of Ag$_{14}$ and its ions, following the
negative-to-neutral-to-positive scheme, serve to illustrate the MDSS method
and demonstrate the potential
of pump-and-probe ultrashort laser pulses as a spectroscopy of cluster
shape vibrations. 

We further remark that the MDSS method can be generalized  
to provide a dynamical theory for metal-cluster fission (see Refs.\ 
[9(a),24$-$26]). 
In analogy with recent dynamical nuclear fission studies,
\cite{carj} such a theory will provide efficient means for calculating 
fission rates, branching ratios between fission channels,  
the dynamics of fission isomers [9(a),25], as well as
the kinetic energy distribution \cite{brec} of the fission fragments for a 
broad range of cluster sizes.

This research is supported by the US Department of Energy (Grant No.
FG05-86ER-45234). Studies were performed at the 
Georgia Institute of Technology Center for Computational Materials Science.

\begin{figure}
\caption{
PES's of Ag$_{14}^-$ (bottom panel), Ag$_{14}$ (middle), and Ag$_{14}^+$
(top). 
$a$ and $b$ denote dimensionless semi-axes of the ellipsoidal 
shapes (the third semi-axis $c=1/ab$; see text for details). 
The contour lines 
correspond to increments of 0.1 eV in the total energy of the clusters.
The Lissajous-type figure
(heavy solid line), inside one of the basins associated with the oblate 
isomer of Ag$_{14}$, portrays one of the 3-ps, constant-energy trajectories
at $T=300$ K (see text).} 
\end{figure}
\begin{figure}
\caption{
vIP surface for Ag$_{14}$. The contour lines correspond to increments of 0.1 
eV. $a$ and $b$ denote dimensionless semi-axes of the ellipsoidal 
shapes (the third semi-axis $c=1/ab$). 
}
\end{figure}
\begin{figure}
\caption{
The time evolution of the probability ${\cal P} (t)$ (see text) for
$T=300$ K (dashed-dotted lines) and $T=600$ K (dotted lines). The 
two-photon energy of the photoionizing probe-laser pulse equals
5.85 eV (lower pair of curves labelled 1 and 2) 
and 5.90 eV (upper pair of curves labelled 3 and 4).
}
\end{figure}


\begin{references}
\bibitem{note11}
In the case of FPMD methods, such collective shape oscillations 
may be analyzed by considering the time variations of the moments of inertia 
of the clusters and/or the eccentricity parameter, which can be calculated
from the positions of the individual ions. For certain small sodium 
clusters this was done in Ref.\ \onlinecite{andr}, where it was found that 
the shape oscillations occur on a time scale of 350 -- 500 fs, 
comparable to the results obtained by us in the present study of Ag$_{14}$.
\bibitem{note2}
In this paper, we use the semiempirical version of the SCM approach.
In the case of Ag$_N$ clusters,
the parameters entering into these calculations are 
$U_0=-0.045$, $r_s=3.01$ a.u., $t=0.47$ a.u., $\delta_0=0$ for the cations,
$\delta_0=1.31$ a.u. for the anions, $\delta_2=0$,
$W=4.26$ eV, $\alpha_v=-8.061$ eV, $\alpha_s=2.05$ eV, and $\alpha_c=0.86$ 
eV (see Ref.\ \onlinecite{yann1} for the meaning of these parameters). 
The mass of the delocalized valence electrons was taken equal to the 
free-electron mass.
For the coinage metals (e.g., Cu, Ag, and Au), contributions
from atomic $d$ electrons enter mainly through the value of the 
work function ($W$), and to a lesser extent through the surface 
tension ($\alpha_s$) and curvature ($\alpha_c$) coefficients,
which are chosen to coincide with the 
experimental values, rather than with those derived through jellium-LDA
calculations (for details, see Ref.\ \onlinecite{yann1}).
\bibitem{note6}
To simulate energy exchange between the dynamical collective shape
variables and the other degrees of freedom of the cluster, one can
include frictional terms in the equations of motion. This would result in
damping of the collective shape dynamics; for the short trajectories used here,
such effects are not expected to have a significant influence.
\bibitem{note4}
In our constant-energy simulations, the total energy 
is conserved within $10^{-7}$ eV. In the case of the Nos\'{e}-Hoover dynamics,
the associated pseudo-energy is conserved within $10^{-6}$ eV.
\end{references}
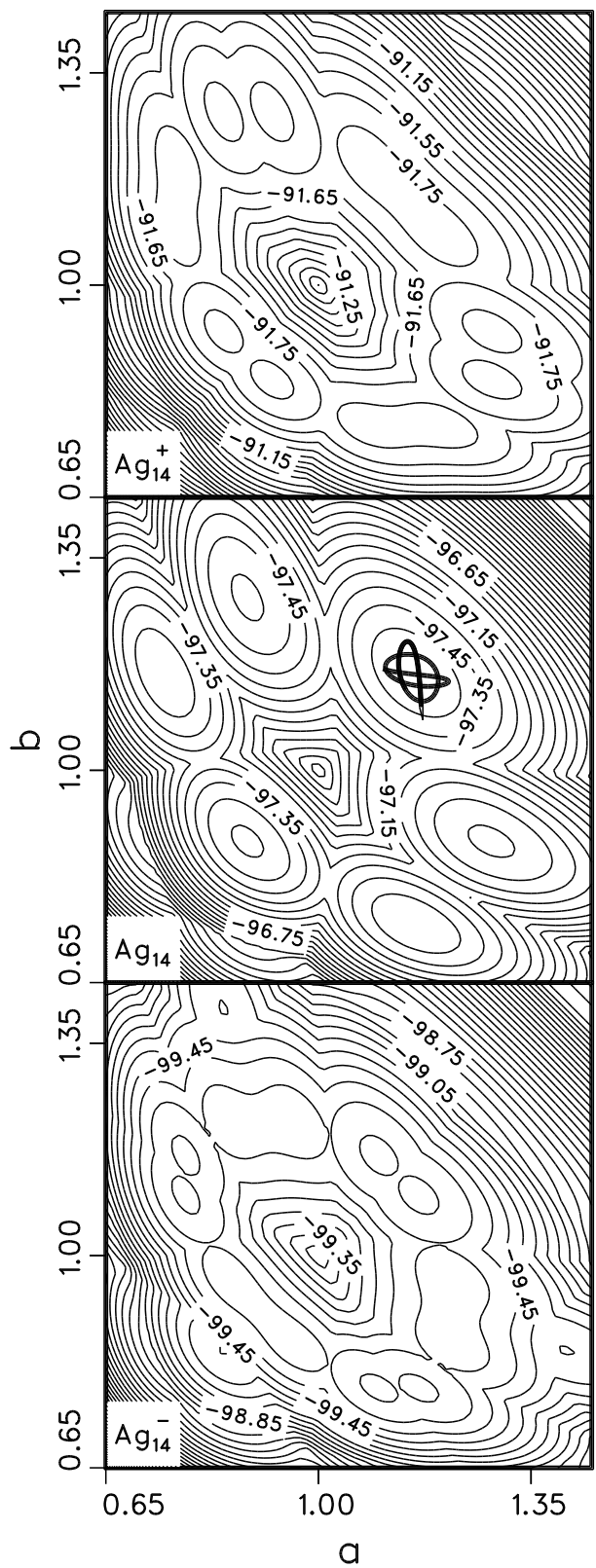
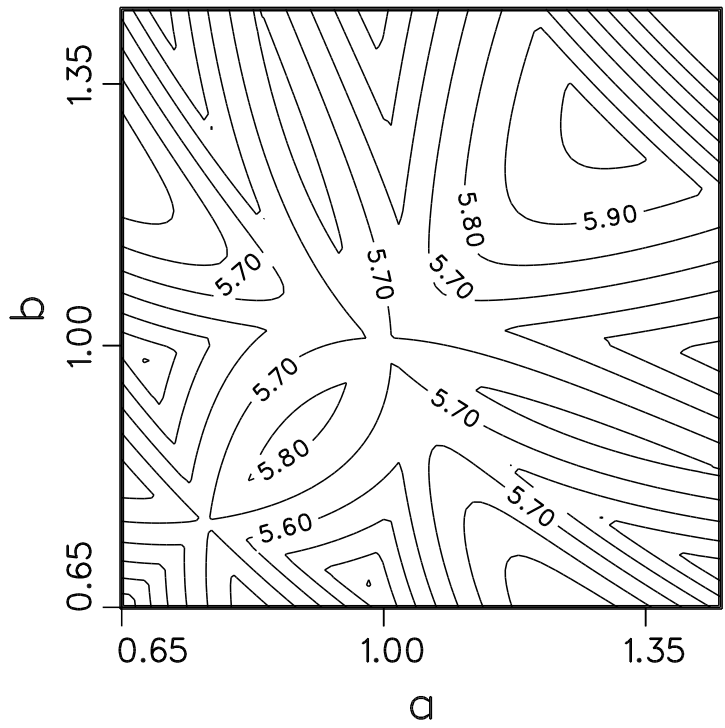
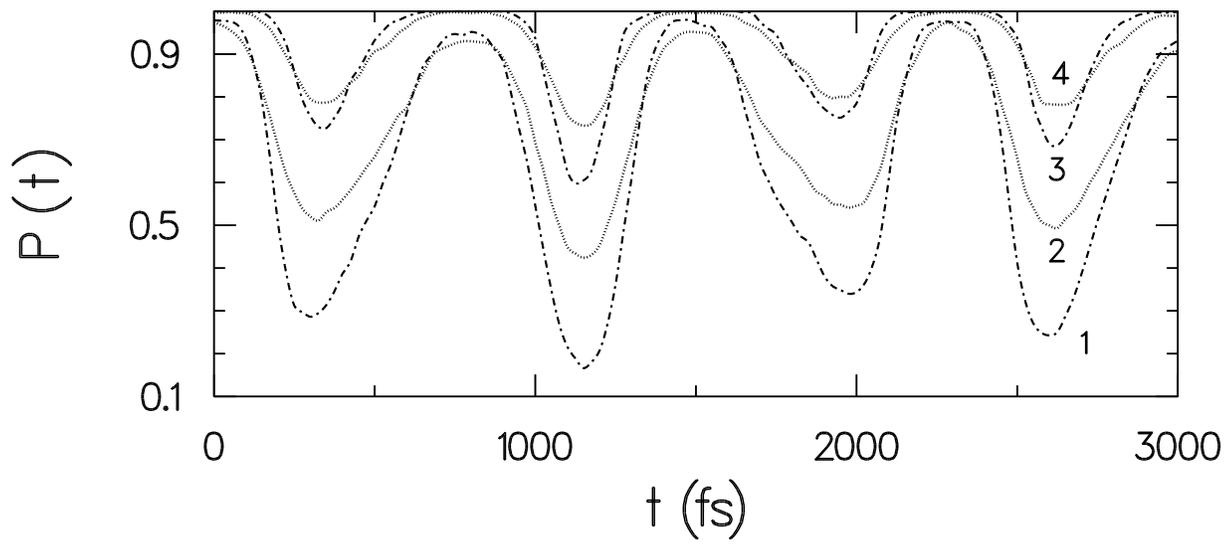
\end{document}